\documentclass{article}
\begin{document}
\begin{center}
WZ COUPLINGS FOR GENERALIZED SIGMA-ORBIFOLD FIXED-POINTS \\ [.25in]
by Juan F. Ospina G.
\end{center}
\begin{center}
ABSTRACT \\ [.25in]
\end{center}
The Wess-Zumino couplings for generalized sigma-orbifold fixed-points  are presented and the generalized GS 6-form that encoding the complete sigma-standard gauge-gravitational-non standard gauge anomaly and its opposite inflow is derived.

\section{Introduction}
The result that this paper presents is about WZ couplings for generalized sigma-orbifold fixed-points. The usual orbifold fixed-points do not involve any kind of  gauge fields, the corresponding GS 6-form that enconding the complete gauge-gravitational anomaly and its opposite inflow only involves the standard gauge field of the D-branes.
From the other side, the sigma-orbifold fixed-points  have one sigma-gauge  field and  the corresponding GS 6-form that enconding the complete sigma-gauge-gravitational anomaly and its opposite inflow involves both the standard gauge field of the D-branes and such sigma-gauge field. 
The generalized sigma-orbifold fixed-points that this paper considers have two gauge fields: the sigma-gauge field and one non-standard SO(2n)-gauge field, and then, the corresponding generalized GS 6-form that encoding the complete sigma-standard gauge-gravitational-non standard gauge anomaly and its opposite inflow, involves three gauge fields: the standard gauge field of the D-branes, the sigma-gauge field and the non-standard SO(2n)-gauge field.  
The aim of the present paper is to display the Wess-Zumino couplings to the
RR forms for such generalized sigma-orbifold fixed-points.

For the usual orbifold fixed-points the Wess-Zumino couplings have the following form, which can be derived both from index theorems and from direct string computation by factorization of the one-loop partition functions: (Scrucca and Serone,hep-th/9912108)  

\begin{center}
{ \mathversion{bold} $ S_{Fk}^{(2k)} = \sqrt{2\pi}\sum_{i_k=1}^{N_k}C_{(2k)}^{i_k}\wedge{Z_{(2k)}}$ }
\end{center}

Where the Mukai vector of RR charges for the usual orbifold fixed-points is given by:

\begin{center}
{ \mathversion{bold} $ Z_{(2k)}(\frac{\rm R}{\rm 4}) =-\frac{\rm 4}{\rm \sqrt{N}}\epsilon_k\sqrt{\frac{\rm \vert{\rm c_k^1c_k^2c_k^3}\vert}{\rm\vert{\rm s_k^1s_k^2s_k^3}\vert }}\sqrt{L(\frac{\rm R}{\rm 4})}  $ }
\end{center}

In these  formulaes C is the vector of the RR potential forms. L is the Hirzebruch genus that generates the Hirzebruch polynomials which are given in terms of Pontryaguin classes for real bundles. The Pontryaguin classes are given in terms of the 2-form curvature of the corresponding real bundle. The formula for Z involves only the gravitational curvature. For each k-twisted sector, s and c are the vectors of sines and cosines respectively of the twist vector v, $\epsilon$ is the sign of the product of the components of the vector  s. $N_k$ is the number of fixed points. Here the relevant group is $Z_N$.

From the other side,for the usual sigma-orbifold fixed-points the Wess-Zumino couplings have the following form, which can be derived both from equivariant index theorems and from direct string computation by factorization of the one-loop partition functions: (Scrucca and Serone,hep-th/0006201)
\begin{center}
{ \mathversion{bold} $ S_{F} = \sqrt{2\pi}\sum_{k=1}^{\frac{\rm 1}{\rm 2}(N-1)}\sum_{i_k=1}^{N_k}\int{C_{(2k)}^{i_k}\wedge{Z_{(2k)}}}$ }
\end{center}     
where,
\begin{center}
{ \mathversion{bold} $ Z_{(2k)}(\frac{\rm R}{\rm 4},\frac{\rm G}{\rm 4}) =-\frac{\rm 4}{\rm \sqrt{N}}\epsilon_k\sqrt{\frac{\rm \vert{\rm C_{2k}}\vert}{\rm\vert{\rm C_k^2}\vert }}\sqrt{L_k(\frac{\rm G}{\rm 4}\epsilon_{2k})}\sqrt{L(\frac{\rm R}{\rm 4})}  $ }
\end{center}
For this case, $L_k$ is the Hirzebruch equivariant factor, G is the sigma gauge field and $C_k$ is the product of the components of the vector s.

In this paper is presented the formula for the WZ couplings  of the generalized sigma-orbifold fixed-points which also have one  aditional SO(2n) Yang-Mills gauge field  .  Such WZ coupling is given by the following formula:
\begin{center}
{ \mathversion{bold} $ S_{F} = \sqrt{2\pi}\sum_{k=1}^{\frac{\rm 1}{\rm 2}(N-1)}\sum_{i_k=1}^{N_k}\int{C_{(2k)}^{i_k}\wedge{Z_{(2k)}}}$ }
\end{center}     
where,
\begin{center}
{ \mathversion{bold} $ Z_{(2k)}(R,G,Y) =-\frac{\rm 4}{\rm \sqrt{N}}\epsilon_k\sqrt{\frac{\rm \vert{\rm C_{2k}}\vert}{\rm\vert{\rm C_k^2}\vert }}\sqrt{\prod_{i=1}^3\frac{\rm sin(pikv_i)}{\rm sin(pikv_i+\frac{\rm G_i}{\rm 4pi})}}\sqrt{A(\frac{\rm R}{\rm 2})\prod_{a=1}^n\frac{\rm cos(piku_a+\frac{\rm Y_a}{\rm 4pi})}{\rm cos(piku_a)}}  $ }
\end{center} 

In this last formula, the productory that involves the sines is the equivariant Dirac-roof factor, the productory that involves the cosines is the equivariant Mayer class, v is the twist vector corresponding to the action of the group $Z_N$ over the normal bundle with respect to space-time of the fixed submanifold, u is the twist vector corresponding to the action of the group $Z_N$ over the SO(2n) bundle on the space-time, Y is the vector of eigenvalues of the 2-form curvature of the SO(2n) bundle, A is the Dirac-roof genus for the tangent bundle of the fixed submanifold.

When the SO(2n) bundle over the space-time is the tangent bundle of the space-time, one obtains the usual formula for Z corresponding to the usual sigma-orbifold fixed-points. For this, the equivariant Mayer class for the tangent bundle of the space-time is factorized as the product of the usual Mayer class for the tangent bundle of the fixed submanifold with the equivariant Mayer factor for the normal bundle of the fixed submanifold. In such case, the vector u is reduced to the vector v and the vector Y is reduced to the sum of the vector of eigenvalues for the 2-form curvature of the tangent bundle of fixed manifold, with the vector G. Finally one can to use the following identity: 

\begin{center}
 
{ \mathversion{bold} $A(\frac{\rm R}{\rm 2})Mayer(\frac{\rm R}{\rm 2}) 
 = L(\frac{\rm R}{\rm 4})$}
\end{center}

In the following section the generalized GS 6-form that enconding the complete sigma-standard gauge-gravitational-non standard gauge anomaly and its opposite inflow, will be given . In the final third section some conclutions are presented.

\section{The generalized GS anomaly-inflow 6-form}

Let E be a SO(2n)-bundle over the space-time and consider a formal factorisation for the total Pontryaguin classs of the real bundle E, which has the following form:

\begin{center}
{ \mathversion{bold} $ p(E) = \prod_{a=1}^n(1+y_a^2)$ }
\end{center}
The total Pontryaguin classs of the real bundle E,has the following formal sumarisation in terms of the corresponding Pontryaguin classes: 
\begin{center}
{ \mathversion{bold} $ p(E) = \sum_{j=0}^{\infty}p_j(E) $ }
\end{center}
The total Mayer class for the real bundle E has the following formal factorisation:
\begin{center}
{ \mathversion{bold} $ Mayer(E) = \prod_{a=1}^ncosh(\frac{\rm y_a}{\rm 2})$ }
\end{center}

The total Mayer class for the real bundle E has the following formal sumarisation in terms of the Mayer polynomials which are formed from the corresponding Pontryaguin classes :
\begin{center}
{ \mathversion{bold} $ Mayer(E) = \sum_{j=0}^{\infty}Mayer_j(p_1(E),...,p_j(E)) $ }
\end{center}
The Mayer polynomials are given by:
\begin{center}
{ \mathversion{bold} $ Mayer_0(p_0(E)) = Mayer_0(1)=1 $ }
\end{center}
\begin{center}
{ \mathversion{bold} $ Mayer_1(p_1(E)) = \frac{\rm p_1(E)}{\rm 8} $ }
\end{center}
\begin{center}
{ \mathversion{bold} $ Mayer_2(p_1(E),p_2(E)) = \frac{\rm p_1(E)^2+4p_2(E)}{\rm 384} $ }
\end{center}
\begin{center}
{ \mathversion{bold} $ Mayer_3(p_1(E),p_2(E),p_3(E)) = \frac{\rm p_1(E)^3+12p_1(E)p_2(E)+48p_3(E)}{\rm 46080} $ }
\end{center}
The pontryaguin classes of the real bundle E have the following realizations in terms of the powers of the 2-form curvature for such bundle.  For this curvature  the y's are the eigenvalues:
\begin{center}
{ \mathversion{bold} $  p_1(E)=p_1(R_E) =-\frac{\rm 1}{\rm 8pi^2}trR_E^2 $ }
\end{center}
\begin{center}
{ \mathversion{bold} $  p_2(E)=p_2(R_E) =\frac{\rm 1}{\rm 16pi^4}[\frac{\rm 1}{\rm 8}(trR_E^2)^2-\frac{\rm 1}{\rm 4}trR_E^4] $ }
\end{center}
\begin{center}
{ \mathversion{bold} $  p_3(E)=p_3(R_E) =\frac{\rm 1}{\rm 64pi^6}[-\frac{\rm 1}{\rm 48}(trR_E^2)^3-\frac{\rm 1}{\rm 6}trR_E^6+\frac{\rm 1}{\rm 8}trR_E^2trR_E^4] $ }
\end{center}
Using all these expretions one can to obtain the following expantion:
\begin{center}
\setlength{\baselineskip}{30pt} 
{ \mathversion{bold} $  Mayer(\frac{\rm R_E}{\rm 2}) = 1+\frac{\rm p_1(R_E)}{\rm 32}+\frac{\rm p_1(R_E)^2+4p_2(R_E)}{\rm 6144}+\frac{\rm p_1(R_E)^3+12p_1(R_E)p_2(R_E)+48p_3(R_E)}{\rm 2949120}+...$ }
\end{center}
Now one has the following expantions:
\begin{center}
\setlength{\baselineskip}{30pt} 
{ \mathversion{bold} $  A(\frac{\rm R}{\rm 2}) = 1-\frac{\rm p_1(R)}{\rm 96}+\frac{\rm 7p_1(R)^2-4p_2(R)}{\rm 92160}+...$ }
\end{center}
\begin{center}
\setlength{\baselineskip}{30pt} 
{ \mathversion{bold} $  L(\frac{\rm R}{\rm 4}) = 1+\frac{\rm p_1(R)}{\rm 48}+\frac{\rm -p_1(R)^2+7p_2(R)}{\rm 11520}+...$ }
\end{center}
Using these three expantions it is easy to obtain the following identities:

\begin{center}
 
{ \mathversion{bold} $A(\frac{\rm R}{\rm 2})Mayer(\frac{\rm R}{\rm 2}) 
 = L(\frac{\rm R}{\rm 4})$}
\end{center} 
\begin{center}
 
{ \mathversion{bold} $A(R)Mayer(R) 
 = L(\frac{\rm R}{\rm 2})$}
\end{center} 
\begin{center}
 
{ \mathversion{bold} $A(2R)Mayer(2R) 
 = L(R)$}
\end{center}
\begin{center}
{ \mathversion{bold} $A(2^qR)Mayer(2^qR) 
 = L(2^{q-1}R)$}
\end{center} 
\begin{center}
 
{ \mathversion{bold} $[A(R)2^kMayer(R)]_{top form} 
 = L(R)_{top form}$}
\end{center}

Now, the group $Z_N$ can be thought to act like the automorphism group of the tangent bundle over space-time, of the standard gauge bundle and of the
SO(2n)-bundle E. For the normal bundle respect to the space-time of the fixed submanifold, the group $Z_N$ acts according to the twist vector v. For the SO(2n)-bundle E over the space-time,  the group $Z_N$ acts according to the twist vector u. Then, the equivariant Mayer class for E has the following factorization:

\begin{center}
{ \mathversion{bold} $ Mayer_k(\frac{\rm y}{\rm 2}) = \prod_{a=1}^n\frac{\rm cosh(ipiku_a+\frac{\rm y_a}{\rm 4})}{\rm cosh(ipiku_a)}$ }
\end{center}

When one defines:

\begin{center}
{ \mathversion{bold} $ y_a = \frac{\rm iY_a}{\rm pi }$ }
\end{center}

one obtains:

\begin{center}
{ \mathversion{bold} $ Mayer_k(\frac{\rm Y}{\rm 2}) = \prod_{a=1}^n\frac{\rm cos(piku_a+\frac{\rm Y_a}{\rm 4pi})}{\rm cos(piku_a)}$ }
\end{center}

When the ten dimentional space-time is of the form $X=R^{1,3}x T^6$ and the fixed submanifold by the action of the group $Z_N$ is $N_k$ copies of $R^{1,3}$   , then the equivariant Dirac-roof factor has the following factorization, where the twist vector v corresponds to the action of the group $Z_N$ over the normal bundle with respect to the space-time of the fixed submanifold and G is the sigma gauge field:

\begin{center}
{ \mathversion{bold} $ A_k(\frac{\rm G}{\rm 2}) = \prod_{i=1}^3\frac{\rm sin(pikv_i+\frac{\rm G_i}{\rm 4pi})}{\rm sin(pikv_i)}$ }
\end{center}

Using all these notations, the Mukay vector of RR charges for the generalized sigma orbifol fixed-points can be writen as follows:

\begin{center}
{ \mathversion{bold} $ Z_{(2k)}(R,G,Y) =-\frac{\rm 4}{\rm \sqrt{N}}\epsilon_k\sqrt{\frac{\rm \vert{\rm C_{2k}}\vert}{\rm\vert{\rm C_k^2}\vert }}\sqrt{A_k(\frac{\rm G}{\rm 2})}\sqrt{A(\frac{\rm R}{\rm 2}) Mayer_k(\frac{\rm Y}{\rm 2})}  $ }
\end{center} 

For the particular case when the SO(2n) gauge bundle E is the SO(10) tangent bundle of the space-time X, the equivariant Mayer class for E has the following factorization, where x is the vector of eigenvalues of the 2-form curvatur R of the tangent bunlde of the fixed submanifold:

\begin{center}
{ \mathversion{bold} $ Mayer_k(\frac{\rm R_X}{\rm 2}) = \prod_{a=1}^5\frac{\rm cos(piku_a+\frac{\rm Y_a}{\rm 4pi})}{\rm cos(piku_a)}=(\prod_{i=1}^3\frac{\rm cos(pikv_i+\frac{\rm G_i}{\rm 4pi})}{\rm cos(pikv_i)})(\prod_{j=1}^2\cos(\frac{\rm ipix_j}{\rm 4pi}))= Mayer_k(\frac{\rm G}{\rm 2})Mayer(\frac{\rm R}{\rm 2})$ }
\end{center}

Using this last factorization it is easy to obtain the usual formula for the Mukay vector of the usual sigma orbifold fixed-points:

\begin{center}
{ \mathversion{bold} $ Z_{(2k)}(R,G,Y) =-\frac{\rm 4}{\rm \sqrt{N}}\epsilon_k\sqrt{\frac{\rm \vert{\rm C_{2k}}\vert}{\rm\vert{\rm C_k^2}\vert }}\sqrt{A_k(\frac{\rm G}{\rm 2})}\sqrt{A(\frac{\rm R}{\rm 2})Mayer_k(\frac{\rm G}{\rm 2})Mayer(\frac{\rm R}{\rm 2}) }=-\frac{\rm 4}{\rm \sqrt{N}}\epsilon_k\sqrt{\frac{\rm \vert{\rm C_{2k}}\vert}{\rm\vert{\rm C_k^2}\vert }}\sqrt{A_k(\frac{\rm G}{\rm 2})Mayer_k(\frac{\rm G}{\rm 2})}\sqrt{A(\frac{\rm R}{\rm 2})Mayer(\frac{\rm R}{\rm 2}) } = -\frac{\rm 4}{\rm \sqrt{N}}\epsilon_k\sqrt{\frac{\rm \vert{\rm C_{2k}}\vert}{\rm\vert{\rm C_k^2}\vert }}\sqrt{L_k(\frac{\rm G}{\rm 4})}\sqrt{L(\frac{\rm R}{\rm 4}) }= Z_{(2k)}(\frac{\rm R}{\rm 4},\frac{\rm G}{\rm 4})$ }
\end{center} 
Here are used the following identities:

\begin{center}
 
{ \mathversion{bold} $A(\frac{\rm R}{\rm 2})Mayer(\frac{\rm R}{\rm 2}) 
 = L(\frac{\rm R}{\rm 4})$}
\end{center}

\begin{center}
 
{ \mathversion{bold} $A_k(\frac{\rm G}{\rm 2})Mayer_k(\frac{\rm G}{\rm 2}) 
 = L_k(\frac{\rm G}{\rm 4})$}
\end{center}

Now, the  total generalized GS couplings can be obtained by summing the D-brane and generalized sigma orbifold fixed-points contributions.  For the D-brane the relevant coupling has the following form (Scrucca and Serone, hep-th/0006201):

\begin{center}
{ \mathversion{bold} $ Y_{(2k)}(R,G,F) =\frac{\rm 1}{\rm \sqrt{N}}\epsilon_{2k}\sqrt{\frac{\rm 1}{\rm\vert{\rm C_{2k}}\vert }}ch_{2k}(\epsilon_{2k}F)\sqrt{A_k(G\epsilon_{k})}\sqrt{A(R) }  $ }
\end{center}

Defining the quantities $X_{(2k)} =Y_{(2k)}+Z_{(2k)}$, one has:
\begin{center}
{ \mathversion{bold} $ S_{GS} = \sqrt{2\pi}\sum_{k=1}^{\frac{\rm 1}{\rm 2}(N-1)}\sum_{i_k=1}^{N_k}\int{C_{(2k)}^{i_k}\wedge{X_{(2k)}}}$ }
\end{center}

Using the explicit forms of the Mukay vectors of RR charges for the D-branes and generalized sigma orbifol fixed-points and the tadpole cancellation condition, one can to check that the total RR charges  $X_{(2k)}^{(0)}$ with respect to the RR 4-forms are zero, and the following results for the total
RR charges $X_{(2k)}^{((2)}$ and $X_{(2k)}^{((4)}$ with respect to the RR 2-forms and the RR 0-forms are found:

\begin{center}
{ \mathversion{bold} $ X_{(2k)}^{(2)} =\frac{\rm 1}{\rm\sqrt{N}2pi N_k^{\frac{\rm 1}{\rm 4}}}[itr({\gamma}_{2k}F)+\frac{\rm 1}{\rm 4}tr({\gamma}_{2k})(\sum_{i=1}^{3}G_itan(pikv_i)+ \sum_{a=1}^{n}Y_atan(piku_a))] $ }
\end{center}  
\begin{center}
\setlength{\baselineskip}{30pt} 
{ \mathversion{bold} $ X_{(2k)}^{(4)} ={\epsilon_{2k}}{\frac{\rm -1}{\rm2\sqrt{N}(2pi)^2 N_k^{\frac{\rm 1}{\rm 4}}}}\{tr({\gamma}_{2k}F^2)-\frac{\rm 1}{\rm 64}tr({\gamma}_{2k})tr(R^2)+itr({\gamma}_{2k}F)\sum_{i=1}^{3}G_icot(2pikv_i)-tr({\gamma}_{2k})[\sum_{i=1}^{3}\frac{\rm 3}{\rm 16}G_i^2{tan^2(pikv_i)}+\sum_{i\ne{j}=1}^{3}\frac{\rm cos(2pikv_i)cos(2pikv_j)-cos^2(pikv_i)cos^2(pikv_j)}{\rm 2sin(2pikv_i)sin(2pikv_j)}G_iG_j+\sum_{a=1}^{n}\frac{\rm Y_a^2(2cos^2(piku_a)+sin^2(piku_a))}{\rm 16cos^2(piku_a)}-\sum_{a=1}^{n}{\sum_{i=1}^{3}\frac{\rm sin(piku_a)cos(pikv_i)}{\rm 8cos(piku_a)sin(pikv_i)}Y_aG_i}-\sum_{a\ne{b}=1}^{n}\frac{\rm sin(piku_a)sin(piku_b)}{\rm 8cos(piku_a)cos(piku_b)}Y_aY_b]\} $ }
\end{center}

These results are generalizations of the equations (6.10) and (6.11) in hep-th/0006201.  When the SO(2n) bundle is the SO(10) tangent bundle, the equations in this paper are reduced to the equations (6.10) and (6.11) in hep-th/0006201.  For such case the vector u is reduced to the vector v and the vector Y is reduced to $G\oplus{(i\pi{x})}$

Finally, one arrives at a very simple factorized expression for the 6-form that are encoding the complete sigma- standard gauge-gravitational-non standard gauge anomaly and its opposite inflow(Scrucca and Serone, hep-th/0006201):

\begin{center}
{ \mathversion{bold} $ A^{(6)} =I^{(6)} = i\sum_{k=1}^{\frac{\rm 1}{\rm 2}(N-1)}N_{k}X_{(2k)}^{(2)}\wedge{X_{(2k)}^{(4)}}$ }
\end{center}

\section{Conclutions}

Using both Mayer class and equivariant Mayer class it is possible to write the WZ couplings for certain generalized sigma orbifold fixed-points.  This involves a new non standard SO(2n) gauge bundle.  When such new bundle is the SO(10) tangent bundle of the ten dimensional space-time of the superstrings theories, then one can to obtain the WZ couplings for the usual 
sigma orbifold fixed-points. Finally when the new WZ coupling for the such generalized sigma orbifold fixed-points is combined with the usual WZ coupling for the usual Dp-brane, on can to obtain the generalized 6 form that are encoding the complete anomaly and its opposite inflow.

\section{References}

\subsection{About WZ couplings  for Dp-branes and Op-planes}
\setlength{\baselineskip}{20pt}
J. Morales, C. Scrucca and M. Serone, Anomalous couplings for D-branes and O-planes, hep-th/9812071

B.Stefanski,Jr., Gravitational Couplings of D-branes and O-planes, hep-th/9812088

K. Dasgupta, D. Jatkar and S. Mukhi, Gravitational couplings and Z2 orientifolds, Nucl. Phys. B523 (1998) 465, hep-th/9707224.

K. Dasgupta and S. Mukhi, Anomaly inflow on orientifold planes, J. High Energy Phys. 3 (1998) 4, hep-th/9709219.

Ben Craps and Frederik Roose, (Non-)Anomalous D-brane and O-plane couplings:the normal bundle,  hep-th/9812149.

Sunil Mukhi and Nemani V. Suryanarayana,  Gravitational Couplings, Orientifolds and M-Planes,  hep-th/9907215

\subsection{About WZ couplings  for generalized Op-planes}
\setlength{\baselineskip}{20pt}

J.F. Ospina, Gravitation couplings for generalized Op-planes, hep-th/0006076,hep-th/0006095, hep-th/0006149

\subsection{About WZ couplings  for sigma orbifold fixed points}
\setlength{\baselineskip}{20pt}
C. Scrucca and M. Serone, Gauge and gravitational anomalies in D=4 N=1 orientifolds, hep-th/9912108
C. Scrucca and M. Serone, hep-th/0006201

\subsection{About Mayer classes}
\setlength{\baselineskip}{20pt}

F. Hirzebruch, Topological Methods in Algebraic Geometry, 1978

Christian Bar,  Elliptic Symbols, december 1995, Math. Nachr. 201, 7-35 (1999)

Keke Li,  Character-valued Index Theorems In Supersymmetric String Theories,
CTP  1460, May 1987

J.F. Ospina, A Heterotic Susy version of  Mayer integrality theorem, hep-th/9606186
   
\end{document}